# Nonlinear Pulse Evolutions in the Laser Wakefield Accelerator through a new Quasi-Static Theory

J. Yazdanpanah

*The Plasma Physics and Fusion Research School, Tehran, Iran*

**ABSTRACT**

Beginning from the set of cold-fluid plus Maxwell equations in the instantaneous, Lorentz-boosted Pulse Co-Moving Frame (PCMF), a new quasi-static theory is developed to describe the nonlinear pulse evolutions due to the wakefield excitation, and is verified through comparison with particle-in-cell (PIC) simulations. According to this theory, the plasma-motion can be treated perturbatively and produces quasi-static wakefield in the PCMF, and the pulse envelope is governed by a form of the Schrodinger equation. The pulse evolutions are characterized by local conservation laws resulted from this equation and subjected to Lorentz transformation into the laboratory frame. In this context, new formulas describing the time-behaviors of group velocity, wake amplitude and carrier frequency are derived and best confirmed by simulation data. The spectral evolutions of the radiation are described based on the properties of the Schrodinger equation, predicting the emergence of a new extra-ordinary dispersion branch with linear relation $\omega \approx ck$ ( $c$ is the light speed) approved by simulations.

## I. INTRODUCTION

Inspired by the rapid and continuous developments in the ultra-intense, ultra-short laser-pulse technology, the Laser Wake-Field Accelerator (LWFA) has received increasing interests over the past two decades (see e.g. Refs. [1, 2] and references therein). In LWFA, the intense laser pulse produces a relativistic, ultra-high phase-velocity electron plasma-wave which is able to self-inject and accelerate electrons from the back-ground plasma up to very high energies, in the form of collimated bunches. The performance of LWFA crucially depends on the phase-velocity (which equals the pulse group-velocity) and the amplitude of the plasma-wave at the injection moment, both of which are dynamic quantities [3] during the propagation, owing to the nonlinear pulse evolutions [3-10]. Therefore, these pulse evolutions are very important to LWFA scenario [3-10] as well as the fundamental theory of light-plasma interactions.

Because a rarefied plasma (mostly applied in LWFA) shows a very weak optical responce, many authors have investigated the laser interaction with this plasma within the approximation of very slow pulse-envelope evolutions [3-18]. This approximation is commonly known as the Qausi-Static (QSA) or the Slow Envelope (SEA) Approximation. For the first time, *Bulanov etal* [4] have generalized this approximation into the regime of strong wakefield excitation which includes LWFA. Later on, many authors have applied and extended this approach, in order to discuss different aspects of radiation and wakefield evolutions in LWFA [3, 5-10, 18]. In one dimension, the main features of pulse evolutions in this regime are the gradual frequency red-shift due to depletion, accompanied by the pulse amplitude growth according to the universal adiabatic law [4]. The amplitude growth then leads to the wakefield amplification in the course of time [3]. More recently, *Schroeder etal* [3] have applied the QSA to calculate the *initial* nonlinear pulse group-velocity and early-time variations of the electron plasma wave phase-velocity.

Briefly speaking, in its existing forms, the QSA is defined by imposing very slow time-variations ($\partial/\partial t \approx 0$) on the radiation inside the Pulse Co-Moving Window (PCMW). This window is not a real Lorentz-boosted frame, but rather a mathematical concept defined by subjecting the governing equations into the algebraic transformation $(x,t) \rightarrow (\xi \equiv x - v_g t, t)$ [3-18]. Here $v_g$ is the *constant, linear* group velocity. As the pulse carrier-frequency is subjected to the heavy, classical Doppler-down-shift in PCMW, the radiation entirely (not only its envelope) is turned quasi-static, resulting in approximate time independence of plasma motion in PCMW.

However, in the case of LWFA, attributed to the currently well-accepted pulse deceleration in the *constant-speed* PCMW [4], the validity of existing QSA formulations is restricted to early times; after a transient time, the pulse attains noticeable propagation-speed and oscillation-frequency in PCMW, therefore, quasi-static solutions for pulse and plasma motion are not longer admitted in this window. One way to overcome this problem is to generalize the existing formalism to incorporate the group velocity variations, but this causes the resulting formalism to become mathematically elaborated and physically intractable. On the other hand, calculation of group-velocity evolution not only is of fundamental interest, but also is practically crucial to LWFA (see e. g. [3, 17-18]). This is while, the previous authors have either ignored the pulse deceleration in their calculations [17-18] or obtained the initial value of the group velocity [3] and the long-term evolution of this quantity has remained quite lacking. In addition to this issue, the important issue of wakefield evolutions [3, 6, 9] is not currently well-understood.

In the present work, we describe pulse evolutions based on applying the quasi-static approximation (slow pulse-envelope variations) in the real Lorentz-boosted Pulse Co-Moving Frame (PCMF). Regarding this approach, we remind that, working inside the PCMF has been previously issued by *McKinstrie and DuBois* [19] to study the parametric instabilities, resulting into a more symmetric formalism. Also, as the plasma appears quasi-static in the PCMF – except for very fine oscillations along the propagation axis, the plasma wave can be regarded as an effective flying-potential for the light pulse, analogous to the developed flying mirror models [20-21]. The dramatic advantages of applying the Lorentz-boosted frames in numerical modeling of beam-interaction systems have been recently demonstrated by *Vay* [22]. Here, we demonstrate the analytical efficiency of working inside the Lorentz-boosted PCMF. As will be shown, in this frame, the plasma motion can be treated perturbatively attributed to its high initial velocity, and the pulse envelope is governed by a form of the standard Schrodinger equation in the presence of the flying potential.

The pulse evolutions are characterized by local conservation laws resulted from the Schrodinger equation and subjected to Lorentz transformation into the laboratory frame. In this context, new formulas describing the time-behaviors of group velocity, wake amplitude and carrier frequency are derived and best confirmed through the Particle-In-Cell (PIC) simulations. The spectral evolutions of the radiation are described based on the properties of the Schrodinger equation, predicting the emergence of a new extra-ordinary dispersion branch with linear relation $\omega \approx ck$ ($c$ is the light speed) approved by simulations.

This paper is organized in different sections. In Sec. II, we briefly outline our basic equations. In Sec. III, we discuss the concept of QSA in the PCMF. In Sec. IV, we describe the nonlinear pulse evolutions. In Sec. V we discuss the wakefield evolutions. In Sec. VI, we describe the spectral evolutions of radiation. Finally, in Sec. VII comparison with the simulation results and discussions are given and the paper is concluded.

## II. GOVERNING EQUATIONS

Much nonlinear aspects of the laser interaction with under-dense plasma may be investigated using the well-known set of cold fluid equations plus Maxwell equations. Here, we apply these equations in the instantaneous Pulse Co-Moving Frame (PCMF). Also, we consider the one dimensional (1D) case, namely a sufficiently broad pulse is assumed such that transverse variations may be neglected. The pulse is p-polarized in $y$ direction and propagates along $x$ direction. The plasma ions are supposed unaffected due to their heavy mass. With these assumptions our basic equations read as follows in the PCMF,

$$\left\{\frac{\partial n_e}{\partial t} + \frac{\partial}{\partial x}[n_e v_{ex}] = 0\right\}\bigg|_{PCMF} \tag{1a}$$

$$\left\{\frac{\partial}{\partial t}[p_{ex} + \frac{e\beta_g}{c}\phi] = -\frac{\partial}{\partial x}[m_e c^2 \gamma_e - e\phi]\right\}\bigg|_{PCMF} \tag{1b}$$

$$\left\{p_{ey} = eA_y\right\}\bigg|_{PCMF} \tag{1c}$$

$$\left\{[\frac{\partial^2}{\partial x^2} - \frac{\beta_g}{c}\frac{\partial}{\partial x \partial t}]\phi = \frac{m_e \gamma_g \omega_p^2}{e}[\frac{n_e}{n_{e0}} - 1]\right\}\bigg|_{PCMF} \tag{1d}$$

$$\left\{[\frac{\partial^2}{\partial x^2} - \frac{1}{c^2}\frac{\partial^2}{\partial t^2}]A_y = \frac{\Omega_p^2}{c^2}A_y\right\}\bigg|_{PCMF} \tag{1e}$$

where $n_e$, $\mathbf{p}_e$, $m_e$, $e$, $\gamma_e = (1-\mathbf{v}_e.\mathbf{v}_e/c^2)^{-1/2}$, $\mathbf{v}_e = \mathbf{p}_e/\gamma_e m_e$, and $\phi$ and $\mathbf{A}$ stand for electron density, momentum, mass, electric charge, relativistic gamma factor, velocity, and scalar and vector potentials, respectively. $\Omega_p^2 \equiv \omega_p^2 n_e/n_{e0}\gamma_e$ where $n_e/\gamma_e$ is the is the proper plasma density and $\omega_p = (n_{e0}e^2/\varepsilon_0 m_e)^{1/2}$ is the usual invariant (based on undisturbed laboratory density) plasma frequency ($n_{e0}$ and $\varepsilon_0$ are respectively the undisturbed plasma density and the vacuum permittivity). $\beta_g \equiv v_g/c$ and $\gamma_g = (1-\beta_g^2)^{-1/2}$. The notation $X|_{PCMF}$ indicates that $X$ is measured in the PCMF. The Coulomb gauge is applied on potentials *in the laboratory frame*, i.e. $A_x = 0$ giving $\{A_x = -\beta_g \phi/c\}|_{PCMF}$.

## III. THE CONCEPT OF QSA IN THE PCMF

### A. DEFINITION AND SINGLE PARTICLE DESCRIPTION

In the linear regime (absence of wake excitation) and in the presence of rarefied plasmas (high group-velocity and weak optical response), the laser pulse appears very simple in its commoving frame (PCMF). Using Lorentz transformations together with the dispersion relation in this regime, it can be easily shown that the carrier frequency and wave-number in the PCMF are given by $\{(\omega_0, ck_0)\}|_{PCMF} = (\omega_p, 0)$. In the other hand, the pulse length is increased according to $L_p|_{PCMF} = \gamma_g L_p$ due to the inverse Fitzgerald contraction. Therefore, the radiation appears as a very long spatial envelope performing very fast temporal oscillations and we can write $\{A_y(x,t) = (1/2)\hat{A}(x)e^{-i\omega_p t} + c.c\}\big|_{PCMF}$ where $\{\partial \hat{A}/\partial x \sim L_p^{-1}\hat{A} \ll \omega_p \hat{A}\}\big|_{PCMF}$.

In the nonlinear regime, on the other hand, the envelope function $\hat{A}$ becomes time-dependent and two different situations are possible, depending on whether we have $\{c^{-1}\partial \hat{A}/\partial t \sim \partial \hat{A}/\partial x\}\big|_{PCMF}$ or $\{c^{-1}\partial \hat{A}/\partial t \ll \partial \hat{A}/\partial x\}\big|_{PCMF}$. In the former case corresponding to long pulses, the radiation is subjected to various parametric-instabilities (see e.g. [16, 19]) which appear convective in the laboratory frame, leading to unsteady propagation in this frame. In the latter case corresponding to short pulses and LWFA, the radiation propagates quasi-steadily in the laboratory frame. This situation defines the quasi-static regime (the scope of QSA) under which the light solution in the *instantaneous* PCMF takes the general form of,

$$\{A_y = (1/2)\hat{A}(x,t)e^{-i\Omega_0 t} + c.c.\}_{PCMF} \tag{2a}$$

Here, $\Omega_0$ is a constant, Lorentz-invariant frequency, analogous to $\omega_p$ in the linear regime and $\hat{A}$ is subjected to the condition,

$$\left\{\frac{\partial \hat{A}}{\partial t} \ll \frac{\partial \hat{A}}{c\partial x} < \Omega_0 \hat{A}\right\}_{PCMF}. \tag{2b}$$

Noting its time independence and using Lorentz transformations, $\Omega_0$ can be written in terms of initial carrier frequency, $\omega_{00}$, and gamma factor, $\gamma_{g0}$, measured in the laboratory frame,

$$\Omega_0 = \frac{\omega_{00}}{\gamma_{g0}}, \qquad (2c)$$

Concerning the oscillatory (time-dependent) form of vector-potential (2a), now the important question is whether the plasma motion is integrable under the action of this potential? –As in the common approach applied *in PCMW*, the QSA is attributed to the heavy, classical Doppler down-shift of the light carrier frequency in this window. The key point to answer this question is that the electron velocity is initially very high in the PCMF and equals the pulse group velocity with the minus sign. Therefore, as long as the laser intensity is not ultra-high (the laser gamma factor $\gamma_L \equiv \sqrt{1+a_0^2}$ is not close to $\gamma_g$), we may apply the method of successive perturbations around the undisturbed (zeroth-order) trajectory to solve the motion equations for electron. In the following we show that this procedure finally results in quasi-static plasma-motion in the PCMF, despite the fast temporal oscillations presented in (2a).

The motion equations (1b) and (1c) are in the Eulerian form and can be easily converted into the usual single-particle (Lagrangian) form using the identity $d/dt = \partial/\partial t + v_{ex}\partial/\partial x$. In this way after some straightforward manipulations we obtain,

$$\left\{\frac{d}{dt}[p_{ex} + \beta_g \frac{e\phi}{c}] = e(1+\beta_g\beta_{ex})\frac{\partial \phi}{\partial x} - \frac{e^2}{2m_e\gamma_e}\frac{\partial A_y^2}{\partial x}\right\}\bigg|_{PCMF} \tag{3a}$$

$$\left\{\frac{d}{dt}[m_e c^2 \gamma_e - e\phi] = -e(1+\beta_g\beta_{ex})\frac{\partial \phi}{\partial t} + \frac{e^2}{2m_e\gamma_e}\frac{\partial A_y^2}{\partial t}\right\}\bigg|_{PCMF} \tag{3b}$$

which are Lorentz-Newton equations in terms of potentials in the PCMF.

We apply the method of successive perturbation to Eqs (3a & 3b) by letting $\{\gamma_e \approx \gamma_g\}|_{PCMF}$ and $\{\beta_{ex} \approx -\beta_g\}|_{PCMF}$ in the right hand sides of these equations in the first approximation. Also, we shale assume a moderate pulse steepening, such that $c^{-1}\partial \hat{A}/\partial x \ll \Omega_0 \hat{A}$. This corresponds to the initial stages of pulse evolution. Now we integrate the resulting equation via the part-by-part rule, making use of identities $A_y^2 = (|\hat{A}|^2/2)(1+\cos[2\Omega_0 t])$ (based on Eq. (2a)), $\partial A_y^2/\partial x = (-1/v_g)(dA_y^2/dt - \partial A_y^2/\partial t)$ and $\partial |\hat{A}|^2/\partial x \approx (-1/v_g) d|\hat{A}|^2/dt$ (based on Eq. (2b)), and ignoring $\partial^2 |\hat{A}|^2/\partial x^2$ throughout the results (based on the moderate steepening). $\phi$ is subjected to same conditions as $\hat{A}$. In this way, after some manipulations, we obtain respectively form Eqs. (3a) and (3b),

$$\left\{ p_{ex}(t) \approx -m_e\gamma_g v_g - (\beta_g + \frac{1}{\beta_g\gamma_g^2})\frac{e\phi(x_e,t)}{c} + \frac{e^2|\hat{A}(x_e,t)|^2}{4m_e\gamma_g v_g} - \frac{e^2}{8m_e\gamma_g\Omega_0}\frac{\partial|\hat{A}|^2}{\partial x}\bigg|_{x=x_e,t}\sin(2\Omega_0 t) \right\}\bigg|_{PCMF}$$

(4a)

$$\left\{ \gamma_e(t) \approx \gamma_g + \frac{e\phi(x_e,t)}{m_e c^2} + \frac{e^2|\hat{A}(x_e,t)|^2}{4m_e^2 c^2\gamma_g}\cos(2\Omega_0 t) - \frac{e^2 v_g}{8m_e^2 c^2\gamma_g\Omega_0}\frac{\partial|\hat{A}|^2}{\partial x}\bigg|_{x=x_e,t}\sin(2\Omega_0 t) \right\}\bigg|_{PCMF}$$

(4b)

where the electron position is taken out of the light region at $t=0$.

In the next order we may use the obtained results (4a) and (4b) in the right hand sides of (3a) and (3b) and integrate the resulting equations again. In this way, we will recover contributions in the form of higher order harmonics of the light which are weaker than the those of the second harmonics given in (4a) and (4b) by a factor of $\gamma_g$.

In Fig.1, we have depicted the analytical results of Eqs. (4a &4b) and compared them with the direct numerical-solution of Eqs. (3a &3b), in the absence of wake excitation ( $\{\phi=0\}|_{PCMF}$ ). The used parameters have been $a_{y0}=1$, $\lambda=1\mu m$, $L_p = c\times 60\text{fs}$, $\gamma_{g0}=10$ (corresponding to typical $n_{e0}/n_{cr}=0.01$), where $\lambda$ and $L_p$ are laser wavelength and pulse length respectively. According to these parameters $\Omega_0$ is computed via Eq. (2c) as $\Omega_0 = \omega_{00}/\gamma_{g0} = 2\pi c/\gamma_{g0}\lambda$. In panels (a, b), we have compared the analytical results (4a) and (4b), which are given in the PCMF, with the direct numerical solutions for the initial position $x_{e0} = 0$, demonstrating excellent agreement. In panel (c), we have plotted $p_{ex}$ versus $\xi$ in the *laboratory frame* both from the direct Lorentz transform of (4a), given in the PCMF, and from the common (commoving window) QSA in the laboratory frame (well-known through previous literatures), observing an excellent agreement between two approaches. The dependence of motion on the initial conditions is shown in panel (d) where plotted are $\beta_{ex} = p_{ex} / \gamma_e$ for two initial positions $x_{e0} = 0$ and $x_{e0} = c\pi / 2\Omega_0$.

## B. THE FLUID DESCRIPTION AND WAKE-FIELD EXCITATION

The single particle solutions (4a) and (4b) can be readily converted to fluid solutions by eliminating $t$ in the right hand side of these equations in terms of the Lagrangian coordinate $x_e = x_{e0} + \int_0^t v_{ex} dt'$. Using the method of successive perturbations discussed above, we get to zeroth order $x_e \approx x_{e0} - v_g t$, hence $t = -(x_e - x_{e0})/v_g$. After this elimination of $t$ in terms of $x$, the time dependence enters into resulting fluid solutions through the initial position $x_{e0}$. In description, at different times, electrons with different initial positions enter into the pulse, seeing the pulse at different initial phases and producing time-dependent plasma-motion. Therefore, the entrance time $t_{en} = x_{e0}/v_g$ may be taken as the time argument of fluid solutions, and the single-particle solutions (e. g. (4a & 4b)) may be converted into fluid solutions by making the transformation

$$\{x_e \to x, \quad t_{en} \to t\}|_{PCMF}. \tag{5}$$

By applying the above transformations, Eq. (4a) and (4b) are respectively transformed to,

$$\left\{p_{ex}(t) \to p_{ex}(x,t) \approx -m_e\gamma_g v_g - (\beta_g + \frac{1}{\beta_g\gamma_g^2})\frac{e\phi}{c} + \frac{e^2|\hat{A}|^2}{4m_e\gamma_g v_g} - \frac{e^2}{8m_e\gamma_g\Omega_0}\frac{\partial|\hat{A}|^2}{\partial x}\sin[2\Omega_0(t-\frac{x}{v_g})]\right\}\Bigg|_{PCMF}$$
,(6a)

$$\left\{\gamma_e(t) \to \gamma_e(x,t) \approx \gamma_g + \frac{e\phi}{m_e c^2} + \frac{e^2|\hat{A}|^2}{4m_e^2 c^2\gamma_g}\cos[2\Omega_0(t-\frac{x}{v_g})] - \frac{e^2 v_g}{8m_e^2 c^2\gamma_g\Omega_0}\frac{\partial|\hat{A}|^2}{\partial x}\sin[2\Omega_0(t-\frac{x}{v_g})]\right\}\Bigg|_{PCMF}$$
(6b)

where $\phi$ and $\hat{A}$ have their own time and space dependence. To the lowest order, Eq. (6b) reproduces the energy conservation law in the PCMF,

$$\{\gamma_e m_e c^2 - e\phi\}|_{PCMF} = \gamma_g m_e c^2. \tag{6c}$$

which after Lorentz transformation, takes the form of $\gamma_e m_e c^2(1-\beta_g\beta_{ex}) - e\phi = m_e c^2$ ($\boldsymbol{\beta}_e \equiv \mathbf{v}_e/c$) in the laboratory frame, just the Hamiltonian constant obtained in the common QSA approach (see e. g. [12]). In the case of transverse-momentum described by Eq. (1c), the transformation (5) gives,

$$\left\{\frac{p_{ey}(t)}{e} = A_y(x_e,t) \rightarrow \frac{p_{ey}(x,t)}{e} = A_y^*(x,t)\right\}\bigg|_{PCMF} \tag{7a}$$

where we used the definition,

$$\left\{A_y^*(x,t) \equiv \frac{\hat{A}}{2}\exp[-i\Omega_0(t-\frac{x}{v_g})] + c.c\right\}\bigg|_{PCMF}. \tag{7b}$$

Concerning this definition, it is worthwhile and useful for our future discussions to obtain its presentation in the laboratory frame. In this regard, we should notice that because $t_{en}$ is measured by an immobile observer in the PCMF, the time argument of (7b), $\{t = t_{en}\}|_{PCMF}$, is the proper-time subjected to the time-delay effect. By other reasoning, as the plasma moves in the PCMF, both the electron-distance from the pulse front and its travel-time to this front are subjected to Fitzgerald contraction. Therefore, the Lorentz transformation for time argument is given by $t|_{PCMF} = t/\gamma$. Using this equation together with Lorentz transformations for position and carrier-frequency, $x|_{PCM} = \gamma_g \xi$, $\omega_0 = \gamma_g \Omega_0$ respectively, and also the Lorentz invariance of transverse momentum, one can write in the laboratory frame,

$$A_y^*(\xi,t) \equiv \frac{\hat{A}(\xi,t)}{2}\exp[i\frac{\omega_0}{v_g}\xi - \frac{\omega_0}{\gamma_g^2}t] + c.c \approx A_y(\xi,t) \tag{7c}$$

which shows that the Lorentz transformation of $A_y^*|_{PCMF}$ is nothing other than vector-potential presentation in the commoving window (PCMW). The marginal difference is only in the slight decrease in the fine wavelength of plasma-motion ($2\pi v_g / \omega_0$) with respect to the laser wave-length ($2\pi v_\varphi / \omega_0 = 2\pi c^2 / \omega_0 v_g$).

The longitude velocity can be calculated via Eqs. (6a & 6b), as,

$$\left\{v_{ex}(x,t) \approx -v_g + \frac{1}{\gamma_g + \frac{e\phi}{m_e c^2}}\left(\frac{-1}{\beta_g \gamma_g^2}\frac{e\phi}{m_e c} + \frac{e^2|\hat{A}|^2}{4m_e^2 c \gamma_g}(1+\cos[2\Omega_0(t-\frac{x}{v_g})]) - \frac{e^2}{4m_e^2 \gamma_g \Omega_0}\frac{\partial|\hat{A}|^2}{\partial x}\sin[2\Omega_0(t-\frac{x}{v_g})]\right)\right\}\bigg|_{PCMF}$$
(8)

where we have used $\gamma_g^{-1}\beta_g^2 = \gamma_g^{-1} - \gamma_g^{-3} \approx \gamma_g^{-1}$ to simplify the result. If we use Eq. (8) into the continuity equation (1a), we find that this equation is decomposed into other two equations respectively for slow and fast components of density perturbation. To the leading order in $\gamma_g^{-1}$ these equations are,

$$\left\{\frac{\partial n_e^{(s)}}{\partial t} + \frac{\partial}{\partial x}[n_e^{(s)} v_{ex}^{(s)}] = 0\right\}\bigg|_{PCMF} \qquad (9a)$$

$$\left\{\frac{\partial n_e^{(2\Omega_0)}}{\partial t} + \frac{\partial}{\partial x}[n_e^{(s)} v_{ex}^{(2\Omega_0)} + n_e^{(2\Omega_0)} v_{ex}^{(s)}] = 0\right\}\bigg|_{PCMF} \qquad (9b)$$

where density is expanded as $n_e \approx n_e^{(s)} + n_e^{(2\Omega_0)}$, and superscripts ($s$) and ($2\Omega_0$) show slow and second-harmonic parts respectively. We may first solve (9a) for $n_e^{(s)}$ using the QSA ($\partial n_e^{(s)} / \partial t \approx 0$), obtaining $\{n_e^{(s)} v_{ex}^{(s)}\}|_{PCMF} = -v_g \gamma_g n_{e0}$ where the relation $n_{e0}|_{PCM} = \gamma_g n_{e0}$ is used. Afterward, we may use the result into (9b) to solve for $n_e^{(2\Omega_0)}$. However, because the space-time dependencies appear in the combination $t - x/v_g$ in this quickly-oscillating component, it will be integrated out through the solution of the Poisson equation. Therefore, it may be either completely discarded or be loosely included in the slow-part through the wakefield calculations. We make the second choice and write,

$$\{n_e v_{ex}\}|_{PCMF} = -v_g \gamma_g n_{e0}, \tag{9c}$$

Using the Lorentz transformations, this equation takes the well-known form $n_e(\beta_g - \beta_{ex}) = n_{e0}\beta_g$ in the laboratory frame (see e. g. [12]).

As according to the above descriptions, no fast oscillations is introduced in $\phi$ through the solution of Poisson equation (1d), therefore this equation is subjected to QSA ($\partial \phi / \partial t|_{PCM} \approx 0$). In addition, if we use Eq. (7b) to write $\gamma = \gamma_L^*(1-\beta_x^2)^{-1/2}$ where $\gamma_L^* \equiv \sqrt{1 + e^2 A_y^{*2} / m_e^2 c^2}$, then substituted this equation into (6c) and then used the result to calculate the density from (9c), (1d) is simplified to,

$$\left\{\frac{\partial^2 \phi}{\partial x^2} = \frac{m_e \gamma_g \omega_p^2}{e}[\frac{\beta_g}{\sqrt{1-\Lambda^2}} - 1]\right\}\bigg|_{PCM}. \tag{10a}$$

where $\Lambda \equiv \gamma_L^* / (\gamma_g + e\phi / m_e c^2)$ is defined. In the limit of ultra-high group velocity, we have $\gamma_g \gg 1$ and $\Lambda \ll 1$, therefore $(1-\Lambda^2)^{-1/2} \approx 1 + \Lambda^2 / 2$ and $\beta_g \approx 1 - 1/2\gamma_g^2$. Using these approximations in (10a), then applying the Lorentz transformations $\phi|_{PCM} = \gamma_g \phi$ and $x|_{PCM} = \gamma_g \xi$ and using Eq. (7c), one easily recovers the well-known equation $\partial^2 \phi / \partial x^2 = (m_e \omega_p^2 / 2e)[\gamma_L^2 (1 + e\phi / m_e c^2)^{-1} - 1]$ in the laboratory where $\gamma_L \equiv \sqrt{1 + e^2 A_y^2 / m_e^2 c^2}$.

Eq. (10a) can be easily cast into a first integral. Defining $\{\psi \equiv 1 + e\phi / \gamma_g m_e c^2\}|_{PCM}$, rewriting Eq. (10a) in terms of this variable, and multiplying the new equation by $\{\partial \psi / \partial x\}|_{PCM}$, we obtain,

$$\left\{ \frac{\partial}{\partial x} \left[ \frac{1}{2} (\frac{\partial \psi}{\partial x})^2 - \frac{\omega_p^2}{c^2} (\frac{\beta_g}{\gamma_g} \sqrt{\gamma_g^2 \psi^2 - \gamma_L^2} - \psi) \right] = \frac{e^2 \omega_p^2}{m_e^2 c^4} \frac{\beta_g}{2\gamma_g \sqrt{\gamma_g^2 \psi^2 - \gamma_L^2}} \frac{\partial A_y^{*2}}{\partial x} \right\}\bigg|_{PCM} \quad (10b)$$

from Eqs. (1c and 6c) it is easily obtained that $\{\gamma_g^2 \psi^2 - \gamma_L^2 = \gamma_e^2 \beta_{xe}^2\}|_{PCM}$ and $\{\psi = \gamma / \gamma_g\}|_{PCM}$. Using these equation together with $\{\partial \psi / \partial x = -eE_x / \gamma_g m_e c^2\}|_{PCM}$ and Eq. (9c), after some straightforward mathematical manipulations, we can rewrite (10b) in the form,

$$\left\{ \frac{\partial}{\partial x} \left[ \frac{1}{2} \varepsilon_0 E_x^2 + n_{e0} m_e c^2 \gamma_e (\beta_g \beta_{xe} + 1) \right] \right\}\bigg|_{PCM} \equiv \frac{1}{\gamma_g} \left\{ \frac{\partial C_w}{\partial \xi} \right\} = \frac{e^2}{2\gamma_g m_e} \left\{ \frac{n_e}{\gamma_e} \frac{\partial A_y^2}{\partial \xi} \right\} \quad (10c)$$

where we have used Lorentz transformations for quantities and defined $C_w = \varepsilon_0 E_x^2 / 2 + n_{e0} m_e c^2 \gamma_e$. With this notation, Eq. (10c) is identically the well-known wake-excitation equation $\partial C_W / \partial \xi = (e^2 n_e / 2\gamma_e m_e) \partial A_y^2 / \partial \xi$ (see e.g. [2, 12]) in the laboratory frame. Behind the pulse, $C_w$ remains constant and equals $\varepsilon_0 E_w^2 / 2$ where $E_w$ is the wake amplitude.

## IV. NONLINEAR PULSE EVOLUTIONS

To calculate the leading order time-evolutions of the laser pulse, we apply the ansatz (2a) with the condition (2c) in the wave equation (1e), and obtain,

$$\left\{ \left[ \frac{\partial^2}{\partial x^2} + \frac{2i\Omega_0}{c^2} \frac{\partial}{\partial t} + \frac{\Omega_0^2 - \Omega_p^2}{c^2} \right] \hat{A} = 0 \right\} \Bigg|_{PCMF} \quad (11)$$

which is exactly the time-dependent Schrodinger equation with well-know properties [23], describing the radiation in the PCMF as a quantum particle confined in the effective potential $V(x) = -(\Omega_0^2 - \Omega_p^2)/c^2$.

Multiplying Eq. (11) by $\hat{A}^*$ and subtracting from the result its complex conjugate, we obtain,

$$\left\{ \Omega_0^2 \frac{\partial}{\partial t} |\hat{A}|^2 + \frac{c^2}{2} \frac{\partial}{\partial x} [i\Omega_0 (\hat{A} \frac{\partial \hat{A}^*}{\partial x} - \hat{A}^* \frac{\partial \hat{A}}{\partial x})] = 0 \right\} \Bigg|_{PCMF}. \quad (12)$$

On the other hand if we take the space-derivative of (11), then multiply the result by $\hat{A}^*$ and then use Eq. (11) in the result, we get,

$$\left\{\frac{-2i\Omega_0}{c^2}\frac{\partial}{\partial t}[\hat{A}^*\frac{\partial \hat{A}}{\partial x}] = \frac{\partial}{\partial x}[\hat{A}^*\frac{\partial^2 \hat{A}}{\partial x^2}] - \frac{\partial}{\partial x}[\frac{\partial \hat{A}}{\partial x}\frac{\partial \hat{A}^*}{\partial x}] + |\hat{A}|^2\frac{\partial V}{\partial x}\right\}\bigg|_{PCMF} \quad (13a)$$

which after substitution of $\partial^2\hat{A}/\partial x^2$ via (11) and summing up the result with its complex conjugate, gives,

$$\left\{\frac{\partial}{\partial t}[i\Omega_0(\hat{A}\frac{\partial \hat{A}^*}{\partial x} - \hat{A}^*\frac{\partial \hat{A}}{\partial x})] + \frac{\partial}{\partial x}[i\Omega_0(\hat{A}^*\frac{\partial \hat{A}}{\partial t} - \hat{A}\frac{\partial \hat{A}^*}{\partial t}) + c^2\frac{\partial \hat{A}}{\partial x}\frac{\partial \hat{A}^*}{\partial x} + \Omega_0^2|\hat{A}|^2]\right\}\bigg|_{PCMF}$$

$$= \left\{\Omega_p^2\frac{\partial}{\partial x}|\hat{A}|^2\right\}\bigg|_{PCMF}$$

(13b)

Within the used approximations and up to the leading order, the electromagnetic energy density, $u_{em} = \varepsilon_0(E_y^2 + c^2 B_z^2)/2$, and momentum density, $g_{em} = \varepsilon_0 E_y B_z$, are respectively given by,

$$\left\{\frac{u_{em}}{\varepsilon_0} = \frac{1}{4}\Omega_0^2|\hat{A}|^2[1 - \cos(2\Omega_0 t - 2\vartheta_0)]\right\}\bigg|_{PCMF}, \quad (14a)$$

$$\left\{\frac{g_{em}}{\varepsilon_0} = \frac{i\Omega_0}{4}[\hat{A}\frac{\partial \hat{A}^*}{\partial x} - \hat{A}^*\frac{\partial \hat{A}}{\partial x}] + \frac{\Omega_0}{2}|\hat{A}|\left|\frac{\partial \hat{A}}{\partial x}\right|\sin(2\Omega_0 t - \vartheta_0 - \vartheta_1)\right\}\bigg|_{PCMF}. \quad (14b)$$

where $\vartheta_0 = \arg(\hat{A})$ and $\vartheta_1 = \arg(\partial\hat{A}/\partial x)$. Apart from a constant factor, the secular parts of the above expressions are identical to those under time-derivative in Eqs. (12) and (13b). Therefore, these equations are nothing other than the local electrodynamics conservation laws (see Jackson [24]) stated in the PCMF – the expression under the space derivative in the left hand side of (13b) is in fact the Maxwell stress.

We integrate Eqs. (12) and (13b) over $x$ to obtain the temporal variations of the secular parts of energy and momentum of the radiation. In this way, after defining the pulse energy and momentum respectively as $\left\{\mathcal{E} = (1/4)\int_{-\infty}^{\infty} \Omega_0^2 |\hat{A}|^2 \, dx\right\}\Big|_{PCMF}$ and $\left\{\mathcal{P}_x = (i\Omega_0/4)\int_{-\infty}^{\infty} [\hat{A}\frac{\partial \hat{A}^*}{\partial x} - \hat{A}^*\frac{\partial \hat{A}}{\partial x}]dx\right\}\Big|_{PCMF}$, we get respectively from (12) and (13b),

$$\mathcal{E}\Big|_{PCMF} = \frac{\Omega_0^2}{4}\gamma_g(t)\int [|\hat{A}|^2(\xi,t)]d\xi = \text{constant} \tag{15a}$$

$$\frac{d\mathcal{P}_x}{dt}\Big|_{PCMF} = -\frac{1}{2}\varepsilon_0 E_w^2 \tag{15b}$$

where in (15a) we have used the identity $x\big|_{PCMF} = \gamma_g \xi$ and in (15b) we have in addition used (10c) in the right hand side of (13b).

Since $\mathcal{P}_x\big|_{PCM} = 0$, a full similarity is recovered between Eqs (15a) and (15b), and the energy-momentum equations of an ordinary relativistic particle in its rest-frame. Here, the total energy in the PCMF $\mathcal{E}_0 \equiv \mathcal{E}\big|_{PCMF}$, which according to (15a) remains time-independent, takes the role of the particle rest-mass. In addition, the energy, $\mathcal{E}$, and the momentum, $\mathcal{P}_x$, form a relativistic four-vector which *irrespective of the chosen reference frame* can be written as [25]

$$(\mathcal{E}, c\mathcal{P}_x) = \mathcal{E}_0(\gamma_g, \gamma_g \beta_g) \quad (16a)$$

in terms of the global group-velocity which accordingly is defined as,

$$v_g = c^2 \frac{\mathcal{P}_x}{\mathcal{E}} \quad (16b)$$

Analogous to the relativistic particle dynamics [25], we can readily apply (15b) in the laboratory frame, i.e. we have $d\mathcal{P}_x/dt = -\varepsilon_0 E_w^2/2$. Making use of this equation together with Eq. (16a) and the identity $d[\gamma_g \beta_g]/dt = \gamma_g^3 d\beta_g/dt$, we obtain

$$\frac{d\beta_g}{dt} = -\frac{1}{2}\frac{\varepsilon_0 c \gamma_{g0} E_w^2}{\gamma_g^3 H_0}. \quad (17a)$$

Here, $H_0 = \gamma_{g0}\mathcal{E}_0$ and $\gamma_{g0} = (1-\beta_{g0}^2)^{-1/2}$ are respectively the initial pulse energy and gamma-factor in which $\beta_{g0}$ is the initial group-velocity. Eq. (17a) derived for the first time in this work fully describes the overall pulse evolutions and deceleration due to wake excitation. At early stages when the wake-amplitude evolutions are ignorable, the solution of Eq. (17a) is obtained in the form,

$$\beta_g = \sin[\tan^{-1}(-\frac{t}{\tau_0} + \gamma_{g0}\beta_{g0})] \quad (17b)$$

where $\tau_0 \equiv 2H_0/\varepsilon_0 c \gamma_{g0} E_w^2$ is the depletion scale-time.

Using Eqs. (17a) and (2c), together with the Lorentz transformations $(\omega_0, k_0) = (\gamma_g \Omega_0, \gamma_g \beta_g \Omega_0)$ (note the carrier wave-number is zero in PCMF due to the zero momentum) we obtain,

$$\omega_0(t) = (\gamma_g / \gamma_{g0})\omega_{00}, \tag{18a}$$

$$k_0(t) = (\gamma_g \beta_g / \gamma_{g0} \beta_{g0}) k_{00} \tag{18b}$$

where $k_{00}$ is the initial carrier wave-number. These equations, also derived for the first time in this paper, give time evolutions of $\omega_0(t)$ and $k_0(t)$ upon the substitution of $\beta_g$ from (17a) or (17b). By combining Eqs. (16a) and (18b) we obtain $\mathcal{E}(t)/\omega_0(t) = \text{constant}$ which is very similar to the adiabatic constant obtained in Ref. [4].

## V. WAKEFIELD EVOLUTIONS

In order to complete our descriptions of pulse evolutions, we need to obtain an expression for time-variations of the wake amplitude. To do this, one way is to numerically calculate this amplitude at each time via putting the instantaneous solution of the Schrodinger equation (11) into the fluid equations in Sec. III.B. However, we have found it more efficient and insightful to use approximate forms of conservation equations (12) and (13b) instead of the full solutions of (11).

As a first approximation, we may neglect the effects of pulse evolutions in expressions under the space-derivative in Eq. (13b) and substitute $\hat{A} \approx \hat{A}_0$ where $\hat{A}_0 \equiv \hat{A}(t=0)$ is the initial pulse envelope. This approximation is strictly valid at early times. Also we may neglect $(\partial \hat{A}/\partial x)(\partial \hat{A}^*/\partial x)$ against $\Omega_0^2 |\hat{A}|^2$ and finally obtain,

$$\left\{\frac{\partial}{\partial t}[i\Omega_0(\hat{A}\frac{\partial \hat{A}^*}{\partial x}-\hat{A}^*\frac{\partial \hat{A}}{\partial x})]=(\Omega_p^2-\Omega_0^2)\frac{\partial}{\partial x}|\hat{A}_0|^2\right\}\bigg|_{PCMF} \tag{19a}$$

Now we perform time-integration and obtain,

$$\left\{i\Omega_0(\hat{A}\frac{\partial \hat{A}^*}{\partial x}-\hat{A}^*\frac{\partial \hat{A}}{\partial x})=(\Omega_p^2-\Omega_0^2)\frac{\partial}{\partial x}|\hat{A}_0|^2 t\right\}\bigg|_{PCMF}=\frac{(\Omega_p^2-\Omega_0^2)}{\gamma_g^2}\frac{\partial}{\partial \xi}|\hat{A}_0|^2 t \tag{19b}$$

where we have used $x|_{PCM}=\gamma_g\xi$ and $dt|_{PCM}=\gamma_g^{-1}dt$ in the right hand side of the result, and the zero initial pulse-momentum is applied. Substituting (19b) into (12) and subjecting the result into the same procedure as applied in obtaining (19b) from (13b), we obtain,

$$|\hat{A}|^2=|\hat{A}_0|^2+\frac{1}{\gamma_{g0}^3}\left(\frac{\omega_p}{2\Omega_0}\right)^2\frac{\partial}{\partial \xi}[\frac{\Omega_{p0}^2-\Omega_0^2}{\omega_p^2}\frac{\partial}{\partial \xi}|\hat{A}_0|^2]c^2 t^2 \tag{20}$$

where $\Omega_{p0}\equiv\Omega_p(t=0)$ is the initial profile of the plasma frequency. Substituting the above expression into the equation $A_y^2(\xi,t)=(|\hat{A}(\xi,t)|^2/2)(1+\cos[2\omega_0\xi/v_g-2\gamma_g^{-2}\omega_0 t])$ (obtained via Eq. (7c)), and then using the result in Eq. (10c) by noting that the oscillatory terms vanish during the integration, we obtain,

$$(\frac{e}{m_e c \omega_p})^2 \left[E_w^2(t) - E_{w0}^2\right] = \frac{c^2 t^2}{\gamma_{g0}^3}\left(\frac{e\omega_p}{2m_e c\Omega_0}\right)^2 \int_0^{-\infty} \frac{n_e}{\gamma_e n_{e0}} \frac{\partial^2}{\partial \xi^2}\left[\frac{\Omega_{p0}^2 - \Omega_0^2}{\omega_p^2} \frac{\partial}{\partial \xi}(A_y^2)_{t=0}\right]d\xi \quad (21)$$

## VI. SPECTRAL EVOLUTIONS OF PULSE

In this section we briefly discuss the spectral evolutions of the radiation in the $k-\omega$ plane (the mode space). The Schrodinger equation (11), is routinely solved via expansion of its solution in terms of its spectral components [23],

$$\left\{\hat{A}(x,t) = \int A_\varpi(x)e^{-i\varpi t}d\varpi\right\}\Big|_{PCMF} \quad (22)$$

where $\varpi$ takes the role of particle energy in quantum mechanics, and is given by $\varpi = \omega|_{PCMF} - \Omega_0$. Since, as is well known in quantum mechanics, for time-independent potentials, *the spectral decomposition of (22) does not change in the course of time* [23], the radiation intensity does not evolves along the $\omega$-axis *in the PCMF*. This phenomenon is already stated in Eq. (15a). Quite contrary, due to incommutability of momentum ($\partial/\partial x$) and energy, the spectrum may evolve along the $k$-axis [23]. Since, the net light-momentum is zero in PCMF, the mean value of spectral-broadening may be stated as $\int k^2 |\hat{A}_k|^2 dk = -\int \hat{A}^*(\partial^2/\partial x^2)\hat{A}dx$ where $\hat{A}_k$ is the Fourier transform of $\hat{A}$. After some mathematical manipulations, it can be found from Eq. (11) that,

$$\left\{\frac{\partial}{\partial t}[\hat{A}^*\frac{\partial^2 \hat{A}}{\partial x^2}+\hat{A}\frac{\partial^2 \hat{A}^*}{\partial x^2}]\right\}\bigg|_{PCMF} = \left\{\frac{c^2}{2i\Omega_0}\frac{\partial}{\partial x}[\frac{\partial \hat{A}^*}{\partial x}\frac{\partial^2 \hat{A}}{\partial x^2}-\frac{\partial \hat{A}}{\partial x}\frac{\partial^2 \hat{A}^*}{\partial x^2}+\hat{A}\frac{\partial^3 \hat{A}^*}{\partial x^3}-\hat{A}^*\frac{\partial^3 \hat{A}}{\partial x^3}]\right\}\bigg|_{PCMF}$$
$$-\left\{\frac{1}{i\Omega_0}[\hat{A}^*\frac{\partial \hat{A}}{\partial x}-\hat{A}\frac{\partial \hat{A}^*}{\partial x}]\frac{\partial}{\partial x}[\Omega_0^2-\Omega_p^2]\right\}\bigg|_{PCMF}$$
(23)

which upon space-integration and making use of (19b) gives the rate of change in the spectral broadening in the PCMF. We do not further discuss this problem, only adding the comment that broadening takes the role of kinetic energy in quantum interpretation –the second term right hand side of (23) is easily recognized as multiplication of velocity and force. Therefore it increases, as the pulse is initially concentrated on the side of the effective potential.

Another important property of spectral evolutions is emergence of new dispersion branches in the laboratory frame which finally remerge into a single branch. In the PCMF, for each $\omega$, the spectrum broadens along the *k*-axis leading to appearance of many new, adjacent dispersion branches parallel to *k*-axis. Since $\omega$-broadening is very small, these branches merge producing the fat-line $\{\omega(k)=\Omega_0\}|_{PCMF}$. Based on Lorentz transformation, in the laboratory frame, this horizontal line appears as an inclined line with relation,

$$\omega = c\beta_g k.$$  (24)

## VII. SIMULATION RESULTS AND DISCUSSIONS

Using a 1D3V (one spatial three velocity dimensions) Particle-In-Cell (PIC) code written by the author [26-27], we have simulated the LWFA scenario over a wide range of physical parameter, all highly supporting the analytical results established above. We exemplify this approval through discussion of two instances of these simulations differing in the pulse amplitude, one with $a_0 = 1$ and the other with $a_0 = 2$, and similar in other used physical and numerical parameters. The laser wavelength and duration are respectively set to $\lambda = 1\mu m$ and $\tau_L = 60[\text{fs}]$, and the pulse shape is initially sinusoidal. The initial plasma profile is step-like with the initial density $n_{e0}/n_c = 0.01$ ($n_c = \varepsilon_0 m_e \omega_0^2 / e^2$ is the critical density), and the initial electron and ion temperatures are set to $k_B T_e / m_e c^2 = 10^{-4} (\sim 50\text{eV})$ and $k_B T_i = 0$ ($k_B$ is Boltzmann constant), respectively. The size of simulation box is 600 λ, with open boundary conditions being applied at its ends for the fields and particles. Each mesh cell is λ/200 long and initially contains 64 macro-particles. Plasma is initialized in the position range $[40\lambda, 960\lambda]$ and the total simulation run time is set as $t = 3\text{ps}$.

In order to examine our results, we should have the initial group velocity, $\beta_{g0}$. Calculation of this quantity has been subject of many extensive studies (see e. g. [17-18]) in the past and more recently has been reexamined by Schroeder etal [3]. We do not summarize the obtained results here as they are elaborated regarding in the case of a general physical state. Instead we use the simple estimate $\gamma_{g0} \approx \sqrt{<\gamma_L>} \omega_{00}/\omega_p$ [17-18] ($\beta_{g0} = \sqrt{1-\gamma_{g0}^{-2}}$) which proven very efficient through our comparison of analytical results with simulations data at different physical conditions. Only when the wakefield amplitude is initially very high, this estimation needs a slight downward adjustment to produce an excellent fit to the simulation results over the full time-period of system evolutions. In the case of our simulation parameters, this estimate gives $\gamma_{g0} \approx 11.1$ for $a_0 = 1$ and $\gamma_{g0} \approx 13.2$ for $a_0 = 2$. The latter is then readjusted to $\gamma_{g0} \approx 12.4$ (0.94% of the original value) as mentioned. Substituting the obtained $\gamma_{g0}$-values into Eq. (2c), $\Omega_0$ is then found as $\Omega_0 = \omega_{00}/\gamma_{g0}$. The obtained initial values are used in following plots of Figs. 2 and 3.

In Fig. 2, we have summarized the main features of pulse evolutions from theory (Eqs. (15a), (17a), (17b) and (18b)) and simulations both for $a_0 = 1$ (left column) and for $a_0 = 2$ (right column). Since the derived formula for time behavior of the wakefield amplitude (Eq. (21)) is not strictly exact, we have found it more convenient, for accuracy-demonstration purpose, to use in the right hand side of Eq. (17a) and its successors the exact wakefield amplitude from simulations. In panels (a, d), for our simulation cases, we have plotted the group velocities versus time both in the presence (direct integration of Eq. (17a)) and the absence (Eq.

(17b)) of wakefield amplification. It is seen that inclusion of the wakefield evolutions is very important.

In order to approve the group velocity formula (17a), for both simulations, we have calculated the instantaneous pulse displacement in the light-speed commoving window through substitution of (17a) into the formula $\Delta\xi(t) = \int_0^t v_g(t')dt' - ct$, and compared the result with the direct-simulation data for the pulse-centroid ($\bar{x}_P = \int xa_y^2 dx / \int a_y^2 dx$) displacement $\Delta\xi(t) = \bar{x}_P(t) - \bar{x}_P(0) - ct$ in Fig.2, panels (b, e). In the case of $a_0 = 2$ (panel (e)) the initial group-velocity adjustment is applied as described above, but the results in the absence of this adjustment are also plotted (dotted line). Generally, it is observed that agreement between the analysis and simulation results is excellent if the used value for the initial group velocity be sufficiently accurate. On the figure, we also presented the instantaneous pulse displacement in the absence of pulse deceleration (dash-dot line) in order to demonstrate the significance of this deceleration.

The accuracy of formulas (15a) and (18b) is demonstrated in Fig. 2, panels (c, f). Here, we have plotted $<a^2>^{-1} = (\int a_y^2 dx)^{-1}$ and $k_0$ versus time respectively from Eq. (15a) and (18b) together with the direct simulation data. Again excellent agreement is observed. We also plotted the formulas in the absence of wakefield amplification to demonstrate the significance of this phenomenon.

In Fig. 3, we have compared the formula (21) for the wakefield evolutions with the simulation results. As this formula is an early-time formula the agreement is not perfect but very good.

In Fig. 4, we have presented the radiation evolutions in the mode-space ($k-\omega$ plane) from simulations and for both laser intensities $a_0=1$ (top) and for $a_0=2$ (bottom). In addition we have plotted the dispersion curves $\omega^2 = \omega_p^2 + c^2 k^2$ (red) and $\omega = ck$ (taupe). It is clearly seen that spectral-broadening increases in the course of time and a new linear dispersion branch is emerged, both verifying the discussions outlined in Sec. VI.

**FIGURE CAPTIONS**

**Figure 1 (Color online): Electron motion inside the laser pulse, as defined in the text: The longitudinal momentum (a) and the gamma factor (b) versus *t* inside the PCMF, and the longitudinal momentum versus ξ in the laboratory frame (c), all drawn both from direct numerical solution (red solid curves) and from analytical expressions (blue, dash-dot curves) and with the initial position *x*$_{e0}$=0 inside PCMF. The longitudinal velocity versus *t* with two different initial positions *x*$_{e0}$=0 (red) and *x*$_{e0}$=k$_p^{-1}$π/2 (taupe) (d).**

**Figure 2 (Color online): Evolution histories of pulse variables for two pulse amplitudes $a_0=1$ (left) and $a_0=2$ (right) calculated both from quoted analytical equations in the presence (solid lines) and the absence (dash-dot lines) of wakefield evolutions, and from PIC simulations (points): Group-velocity (a, d), displacement in the light-speed commoving window (b, e) and the peak-mode wave-number (c, f). In panel (e) the dotted**

curve shows the analytical result without adjustment of the initial gamma factor. In panels ( c) and (f), in addition to the peak-mode wave-number, simulation results are given for $<a^2>^{-1}$ **according to descriptions in the text.**

**Figure 3 (Color online): Evolution histories of wakefield amplitude for two pulse amplitudes** $a_0 = 1$ **(a) and** $a_0 = 2$ **(b) calculated both from Eq. (21) (dash-dot curve) and from PIC simulations (solid curve). In panel (b) the dotted curve shows the analytical result without adjustment of the initial gamma factor.**

**Figure 4 (Color online): Intensity map of radiation in *k*-ω plan at different times quoted on the figures, for two pulse amplitudes** $a_0 = 1$ **(a-c) and** $a_0 = 2$ **(d-f). The dash-dot curves show the dispersion relations** $\omega^2 = \omega_p^2 + c^2 k^2$ **(red) and** $\omega = ck$ **(taupe).**

FIG. 1:

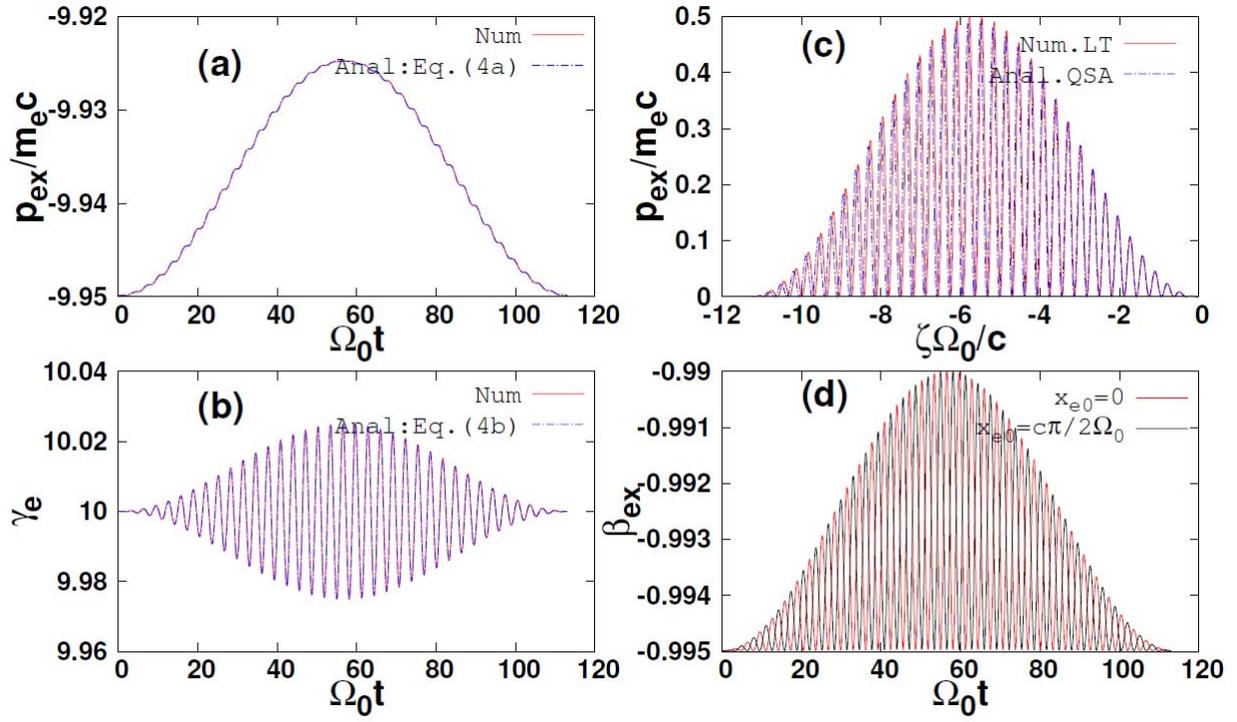

FIG. 2:

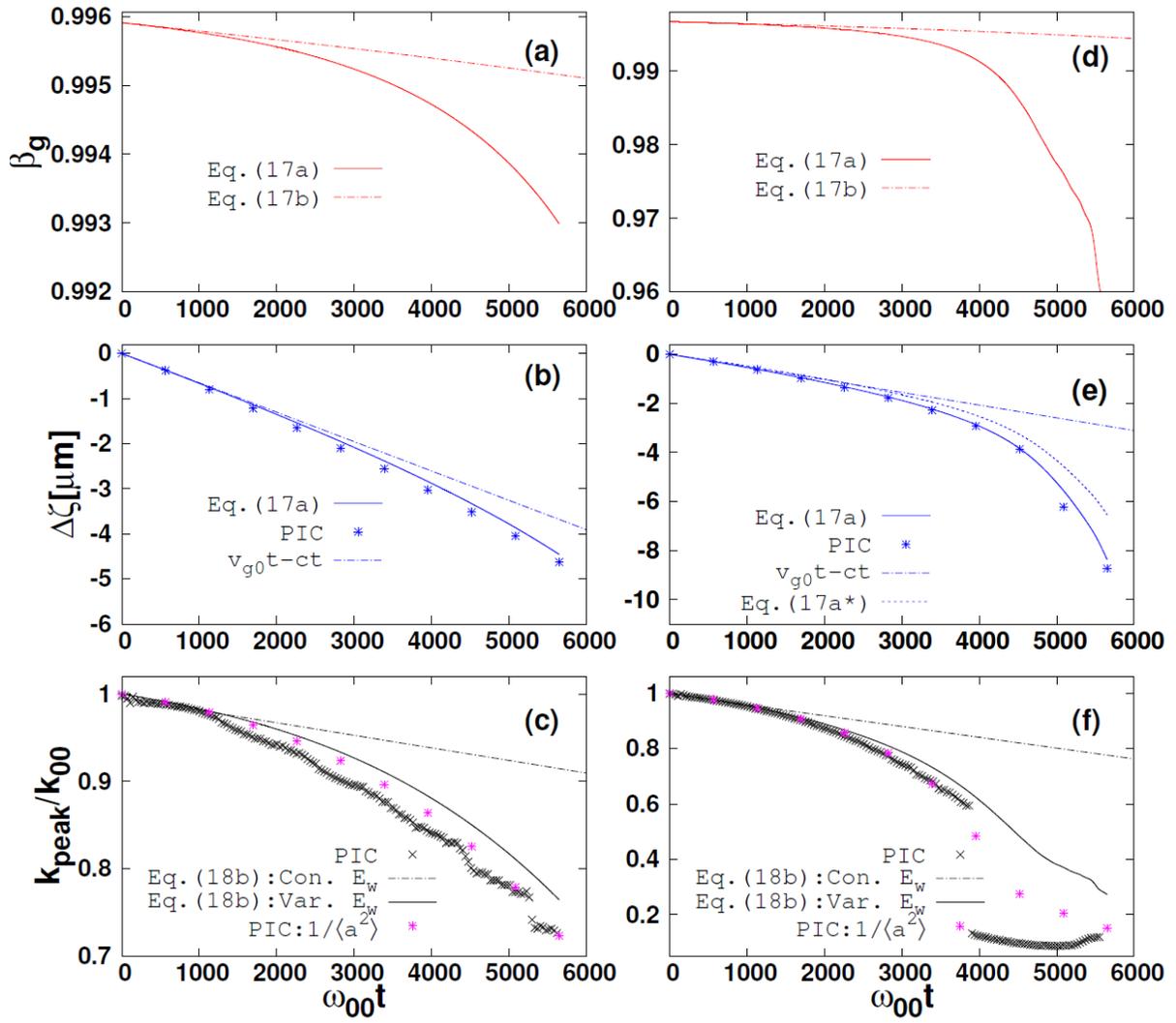

FIG. 3:

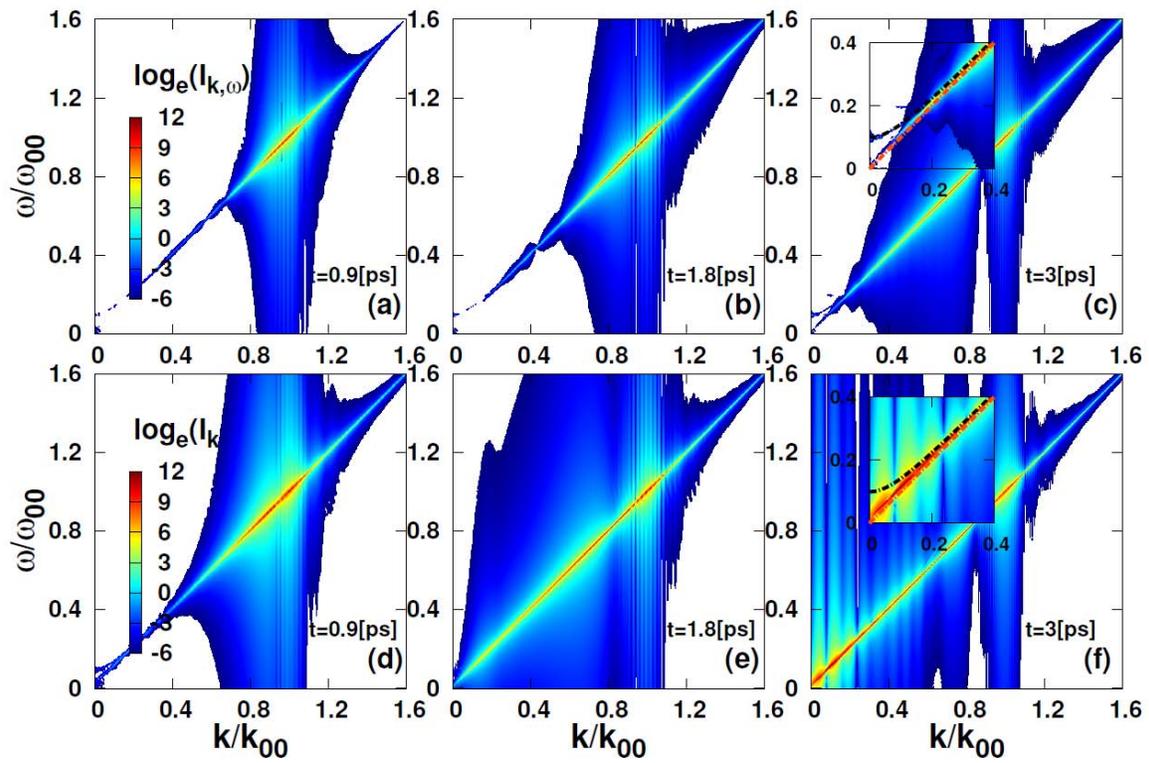

FIG. 4:

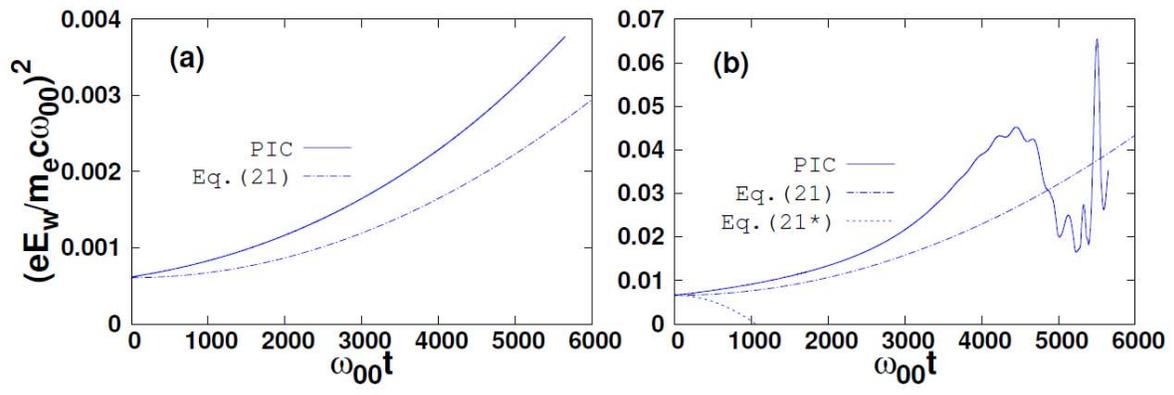